\documentclass[aps,onecolumn,amsmath,amssymb,prl,superscriptaddress,preprint]{revtex4}
\usepackage{newtxtext,newtxmath}
\usepackage{graphicx}
\usepackage{dcolumn}
\usepackage{bm}
\usepackage{color}

\begin{document}
\title{Thickness dependence of spin Peltier effect visualized by thermal imaging technique}
\author{Shunsuke Daimon}
\email{daimon@ap.t.u-tokyo.ac.jp}
\affiliation{Institute for Materials Research, Tohoku University, Sendai 980-8577, Japan}
\affiliation{Advanced Institute for Materials Research, Tohoku University, Sendai 980-8577, Japan}
\affiliation{Department of Applied Physics, The University of Tokyo, Tokyo 113-8656, Japan}
\author{Ken-ichi Uchida}
\email{UCHIDA.Kenichi@nims.go.jp}
\affiliation{Institute for Materials Research, Tohoku University, Sendai 980-8577, Japan}
\affiliation{National Institute for Materials Science, Tsukuba 305-0047, Japan}
\affiliation{Department of Mechanical Engineering, The University of Tokyo, Tokyo 113-8656, Japan}
\affiliation{Center for Spintronics Research Network, Tohoku University, Sendai 980-8577, Japan}
\author{Naomi Ujiie}
\affiliation{ALPS ALPINE CO., LTD., Tokyo 145-8501, Japan}
\author{Yasuyuki Hattori}
\affiliation{ALPS ALPINE CO., LTD., Tokyo 145-8501, Japan}
\author{Rei Tsuboi}
\affiliation{Institute for Materials Research, Tohoku University, Sendai 980-8577, Japan}
\author{Eiji Saitoh}
\affiliation{Institute for Materials Research, Tohoku University, Sendai 980-8577, Japan}
\affiliation{Advanced Institute for Materials Research, Tohoku University, Sendai 980-8577, Japan}
\affiliation{Department of Applied Physics, The University of Tokyo, Tokyo 113-8656, Japan}
\affiliation{Center for Spintronics Research Network, Tohoku University, Sendai 980-8577, Japan}
\affiliation{Advanced Science Research Center, Japan Atomic Energy Agency, Tokai 319-1195, Japan}
\begin{abstract}
Magnon propagation length in a ferrimagnetic insulator yttrium iron garnet (YIG) has been investigated by measuring and analyzing the YIG-thickness $t_{\rm{YIG}}$ dependence of the spin Peltier effect (SPE) in a Pt/YIG junction system. By means of the lock-in thermography technique, we measured the spatial distribution of the SPE-induced temperature modulation in the Pt/YIG system with the $t_{\rm{YIG}}$ gradation, allowing us to obtain the accurate $t_{\rm{YIG}}$ dependence of SPE with high $t_{\rm{YIG}}$ resolution. Based on the $t_{\rm{YIG}}$ dependence of SPE, we verified the applicability of several phenomenological models to estimate the magnon diffusion length in YIG.
\end{abstract}
%
%
%
\maketitle

%
%
%
%
%
Interconversion between spin and heat currents has been extensively studied in the field of spin caloritronics \cite{spincaloritronics1,spincaloritronics2}.
One of the spin-caloritronic phenomena is the spin Seebeck effect (SSE) \cite{SSE3}, which generates a spin current as a result of a heat current in metal/magnetic-insulator junction systems. The Onsager reciprocal of SSE is the spin Peltier effect (SPE) \cite{SPE1,SPE2,SPE3,SPE6,SPEtheory3,SPE8,SPE9,Yamazaki}.
A typical system used for studying SPE and SSE is paramagnetic metal Pt/ferrimagnetic insulator yttrium iron garnet (YIG) junction systems \cite{SPE1,SPE2,SPE3,SSE3,SSE8,SSE20,SSE22,SPE4}, where the spin and heat currents are carried by electron spins in Pt and magnons in YIG \cite{SSEtheory5,SSEtheory11,SPEtheory2,SPEtheory4}.\par
%
%
%
%
%
SPE and SSE are characterized by their length scale including the magnon-spin diffusion length $l_{\rm{m}}$ and the magnon-phonon thermalization length $l_{\rm{mp}}$ in YIG \cite{Duine,SSE36,SPEtheory2}. These length parameters have been investigated by measuring discrete YIG-thickness $t_{\rm{YIG}}$ dependence of SPE and SSE in several Pt/YIG junction systems with different $t_{\rm{YIG}}$ \cite{SSE20,SSE22,SSE26,SSE36,SPE3}. However, different junctions may have different magnetic properties, surface roughness, crystallinity, and interface condition. These variations make it hard to analyze the fine $t_{\rm{YIG}}$ dependence and obtain correct values of the length parameters.\par
In this letter, we measured the $t_{\rm{YIG}}$ dependence of SPE by using a single Pt/YIG system with a $t_{\rm{YIG}}$ gradient ($\nabla t_{\rm{YIG}}$). Since SPE induces temperature change reflecting the local $t_{\rm{YIG}}$ value, we can extract the $t_{\rm{YIG}}$ dependence of SPE from a temperature distribution. The spatial distribution of the SPE-induced temperature change was visualized by means of the lock-in thermography (LIT) \cite{SPE2}. The LIT method allows us to obtain the accurate $t_{\rm{YIG}}$ dependence of SPE with high $t_{\rm{YIG}}$ resolution. By means of the thermoelectric imaging technique based on laser heating \cite{Iguchi2019}, we also measured the $t_{\rm{YIG}}$ dependence of SSE in a single Pt/YIG system, which shows the same behavior as that of SPE. By analyzing the measured $t_{\rm{YIG}}$ dependence of SPE and using phenomenological models, we estimated $l_{\rm{m}}$ and determined the upper limit of $l_{\rm{mp}}$ for YIG. \par
%
%
%
%
%
\begin{figure}[t]
\begin{center}
\includegraphics{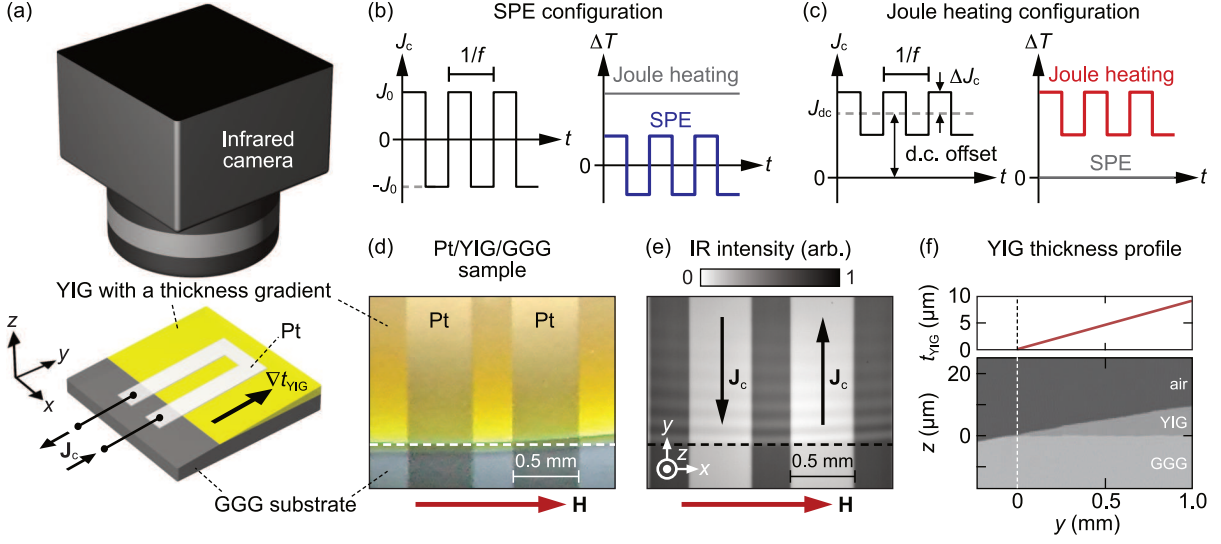}
\caption{(a) Schematic of the SPE measurement using a Pt/YIG/GGG sample by means of the lock-in thermography method. A charge current ${\bf{J}}_{\rm{c}}$ is applied to the U-shaped Pt film fabricated on the YIG film with a thickness gradient $\nabla t_{\rm{YIG}}$.  (b),(c) Time $t$ profile of an input a.c. charge current $J_{\rm{c}}$ and output temperature change $\Delta T$ for the (b) SPE and (c) Joule heating configurations. (d) An optical microscope image of the sample. The yellow (gray) area above (below) the white dotted line corresponds to the YIG film with $\nabla t_{\rm{YIG}}$ (GGG substrate). ${\bf{H}}$ is an applied magnetic field. (e) An infrared image of the sample. (f) A $t_{\rm{YIG}}$ profile and cross-sectional image of the sample obtained with a scanning electron microscope.}
\label{figure1}
\end{center}
\end{figure}
%
%
%
%
%
%
%
%
%
%
The sample used for measuring SPE consists of a Pt film and an YIG film with $\nabla t_{\rm{YIG}}$ [Fig. \ref{figure1}(a),(d)]. $\nabla t_{\rm{YIG}}$ was introduced by obliquely polishing a single-crystalline YIG (111) film grown by a liquid phase epitaxy method on a single-crystalline Gd$_3$Ga$_5$O$_{12}$ (GGG) (111) substrate. The obtained $\nabla t_{\rm{YIG}}$ is almost uniform in the measurement range, which was observed to be $9.2$ $\mu$m per a $1$ mm lateral length by a cross-sectional scanning electron microscopy [Fig. \ref{figure1}(f)]. After the polishing, a U-shaped Pt film with a thickness of 5 nm and width of 0.5 mm was sputtered on the surface of the YIG film. The longer lines of the U-shaped Pt film were along the $\nabla t_{\rm{YIG}}$ direction. In the microscope image of the sample in Fig. \ref{figure1}(d), the yellow (gray) area above (below) the white dotted line corresponds to the YIG film with $\nabla t_{\rm{YIG}}$ (GGG substrate).\par
SPE induces temperature modulation in the Pt/YIG/GGG sample in response to a charge current in the Pt film. When we apply a charge current ${\bf{J}}_{\rm{c}}$ to the Pt film as shown in Fig. \ref{figure1}(a), a spin current is generated by the spin Hall effect in Pt \cite{SHE5,SHE6}.
The spin current induces a heat current across the Pt/YIG interface via SPE. The heat current results in a temperature change $\Delta T$ which satisfies the following relation \cite{SPE1,SPE2}: $\Delta T\propto\left({\bf{J}}_{\rm{c}}\times{\bf{M}}\right)\cdot {\bf{n}}$, where ${\bf{M}}$ and ${\bf{n}}$ are the magnetization vector of YIG and the normal vector of the Pt/YIG interface plane, respectively. Significantly, the SPE-induced temperature change reflects local $t_{\rm{YIG}}$ information because the temperature change induced by SPE is localized owing to the formation of dipolar heat sources \cite{SPE2,SPE3}. Based on the $\nabla t_{\rm{YIG}}$ value and the spatial resolution of our LIT system, we obtained the high $t_{\rm{YIG}}$ resolution of $92$ nm.\par
The procedure of the LIT-based SPE measurements are as follows \cite{SPE2,LIT1,LIT2,LIT3,LIT4,LIT5,AMPE}. To excite SPE, we applied a rectangular a.c. charge current ${\bf{J}}_{\rm{c}}$ with the amplitude $J_0$, frequency $f$, and zero offset to the Pt film [Fig. \ref{figure1}(b)]. By extracting the first harmonic response of a temperature change $\Delta T_{1f}$ in this condition (SPE configuration), we can detect the pure SPE signal free from a Joule heating contribution \cite{SPE2,SPE3}. Here, $\Delta T_{1f}$ is defined as the temperature change oscillating in the same phase as ${\bf{J}}_{\rm{c}}$ because $\Delta T$ generated by SPE follows the ${\bf{J}}_{\rm{c}}$ oscillation [Fig. \ref{figure1}(b)] and the out-of-phase signal is negligibly small \cite{SPE2,SPE3,SPE5,SPE7}. In the LIT measurement, we detect the first harmonic component of the infrared light emission $\Delta I_{1f}$ caused by $\Delta T_{1f}$, where $\Delta I_{1f}=A\cos \phi$ with $A$ and $\phi$ respectively being the lock-in amplitude and phase. We converted $\Delta I_{1f}$ into $\Delta T_{1f}$ by considering spatial distribution of an infrared emissivity $\epsilon$ of the sample \cite{SPE2,SPE3}.
All measurements of SPE were performed at room temperature and atmospheric pressure under a magnetic field with a magnitude of $20$ mT, where ${\bf{M}}$ aligns along the field direction.\par
%
%
%
%
%
\begin{figure}[t]
\begin{center}
\includegraphics{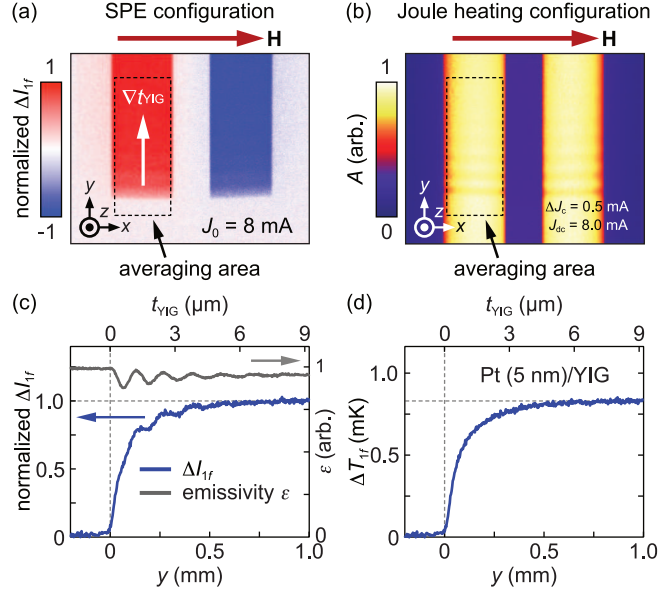}
\caption{(a) Lock-in image of an infrared light emission $\Delta I_{1f}$ induced by SPE. (b) Lock-in amplitude of the infrared emission $A$ induced by the Joule heating. (c),(d) YIG-thickness $t_{\rm{YIG}}$ dependence of (c) $\Delta I_{1f}$ and the emissivity $\epsilon$ and (d) the temperature change $\Delta T_{1f}$ induced by SPE.}
\label{figure2}
\end{center}
\end{figure}
%
%
%
%
%
%
%
%
%
%
%
%
%
%
%
%
%
%
%
Figure \ref{figure2}(a) shows the $\Delta I_{1f}$ image from the Pt/YIG/GGG sample in the SPE configuration with $J_0=8$ mA and $f=5$ Hz. Clear signals appear only on the Pt film with finite $t_{\rm{YIG}}$ but disappears on the Pt/GGG interface with $t_{\rm{YIG}}=0$ [compare Figs. \ref{figure1}(d) and \ref{figure2}(a)]. The sign of $\Delta I_{1f}$ is reversed when ${\bf{J}}_{\rm{c}}$ is reversed, which confirms that the observed infrared signal comes from SPE \cite{SPE1,SPE2}. The spatial profile of $\Delta I_{1f}$ along the $y$ direction is plotted in Fig. \ref{figure2}(c), where the $\Delta I_{1f}$ values are averaged along the $x$ direction in the area surrounded by the dotted line in Fig. \ref{figure2}(a). $\Delta I_{1f}$ gradually increases with small oscillation with increasing $t_{\rm{YIG}}$. The oscillation originates from the oscillation of $\epsilon$ of the sample due to multiple reflection and interference of the infrared light in the YIG film [see the infrared light image of the sample in Fig. \ref{figure1}(e)] \cite{SPE3}. To calibrate the $\epsilon$ oscillation in the $t_{\rm{YIG}}$ dependence of SPE, the infrared emission induced by the Joule heating was measured [Fig. \ref{figure2}(b)], where ${\bf{J}}_{\rm{c}}$ with the amplitude $\Delta J_{\rm{c}}=0.5$ mA, frequency $f=5$ Hz, and d.c. offset $J_{\rm{dc}}=8.0$ mA was used as an input for the LIT measurement [Fig. \ref{figure1}(c)] \cite{SPE2}. Since the temperature change induced by the Joule heating is uniform on the Pt film, the lock-in amplitude of the infrared emission $A$ on the Pt film depends only on the $\epsilon$ distribution. We found that the obtained $t_{\rm{YIG}}$ dependence of $\epsilon$ ($\propto A$ due to the Joule heating) and the $\Delta I_{1f}$ signal due to SPE exhibit the similar oscillating behavior [Fig. \ref{figure2}(c)]. By calibrating $\Delta I_{1f}$ by $\epsilon$, we obtained the $t_{\rm{YIG}}$ dependence of $\Delta T_{1f}$ induced by SPE [Fig. \ref{figure2}(d)]. $\Delta T_{1f}$ monotonically increases with increasing $t_{\rm{YIG}}$.
The saturated $\Delta T_{1f}$ value for $t_{\rm{YIG}}>6$ $\mu$m was determined by using the same Pt/YIG/GGG sample coated with a black-ink infrared emission layer with high emissivity larger than 0.95.\par
\begin{figure}[t]
\begin{center}
\includegraphics{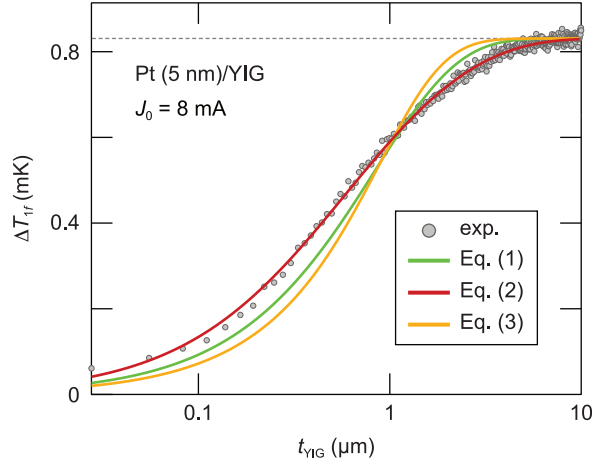}
\caption{Experimental results of the $t_{\rm{YIG}}$ dependence of $\Delta T_{1f}$ induced by SPE for the Pt/YIG/GGG sample and fitting curves using Eqs. (\ref{exponential})-(\ref{Basso}).}
\label{figure3}
\end{center}
\end{figure}

%
%
%
%
%
%
%
%
%
%
The accurate $t_{\rm{YIG}}$ dependence of SPE with high $t_{\rm{YIG}}$ resolution allows us to verify the applicability of several phenomenological models used for discussing the behaviors of SPE. First of all, we found the obtained $t_{\rm{YIG}}$ dependence of the SPE-induced temperature modulation cannot be explained by a simple exponential approximation \cite{SSE20,SSE22,SSE26,SPE3}. Based on the assumption that the magnon diffuses in YIG with the magnon diffusion length $l_{\rm{m}}$, the simple exponential decay has been used for the analysis of the $t_{\rm{YIG}}$ dependence:
\begin{equation}
\label{exponential}
\Delta T\propto 1-{\rm{exp}}\left(-t_{\rm{YIG}}/l_{\rm{m}}\right).
\end{equation}
However, in general, this expression cannot be used for the small thickness region since the exponential function should be modulated by the boundary conditions for the spin and heat currents. In fact, the fitting result using Eq. (\ref{exponential}) significantly deviates from the experimental data when $t_{\rm{YIG}} < 4$ $\mu$m (see the green curve in Fig. \ref{figure3}). The observed continuous $t_{\rm{YIG}}$ dependence of SPE thus requires advanced understanding of the spin-heat conversion phenomena.\par
%
%
%
%
%
%
%
%
%
%
Next we focus on two phenomenological models proposed in Refs. \onlinecite{SPEtheory4} and \onlinecite{SPEtheory1}. The model in Ref. \onlinecite{SPEtheory4} is based on the linear Boltzmann's theory for magnons and the $t_{\rm{YIG}}$ dependence of SPE is described as
\begin{equation}
\label{Cornelissen}
\Delta T \propto \frac{1}{S\coth\left(t_{\rm{YIG}}/l_{\rm{m}}\right)+1},
\end{equation}
where $S$ is a $t_{\rm{YIG}}$-independent constant used as a fitting parameter in our analysis. The red curve in Fig. \ref{figure3} shows the fitting result based on Eq. (\ref{Cornelissen}). We found that Eq. (\ref{Cornelissen}) shows the best agreement with the experimental result and $l_{\rm{m}}$ is estimated to be $3.9$ $\mu$m. We also analyzed the experimental results by the model based on the non-equilibrium thermodynamics \cite{SPEtheory1}:
\begin{equation}
\label{Basso}
\Delta T \propto \frac{\cosh\left(t_{\rm{YIG}}/l_{\rm{m}}\right)-1}{\sinh\left(t_{\rm{YIG}}/l_{\rm{m}}\right)+r\cosh\left(t_{\rm{YIG}}/l_{\rm{m}}\right)},
\end{equation}
where $r$ is a $t_{\rm{YIG}}$-independent constant used as a fitting parameter in our analysis. In contrast to Eq. (\ref{Cornelissen}), Eq. (\ref{Basso}) is less consistent with the experimental results in the whole thickness range (see the yellow curve in Fig. \ref{figure3}) and gives a shorter magnon diffusion length of $0.6$ $\mu$m. From the fitting results using Eqs. (\ref{exponential}-\ref{Basso}), we conclude that Eq. (\ref{Cornelissen}) is the best phenomenological model to explain the experimental results on SPE.\par
To check the reciprocity between SPE and SSE \cite{reciprocity}, we also measured the $t_{\rm{YIG}}$ dependence of SSE by using a single Pt/YIG/GGG sample with $\nabla t_{\rm{YIG}}$. The YIG film used in the SSE measurement was obtained from the same YIG/GGG substrate and the SSE sample was prepared by the same method as that for the SPE sample.
The $\nabla t_{\rm{YIG}}$ value of the SSE sample was determined to be $12.2$ $\mu$m per a $1$ mm lateral length.
%
%
%
%
%
%
%
\begin{figure*}[t]
\begin{center}
\includegraphics{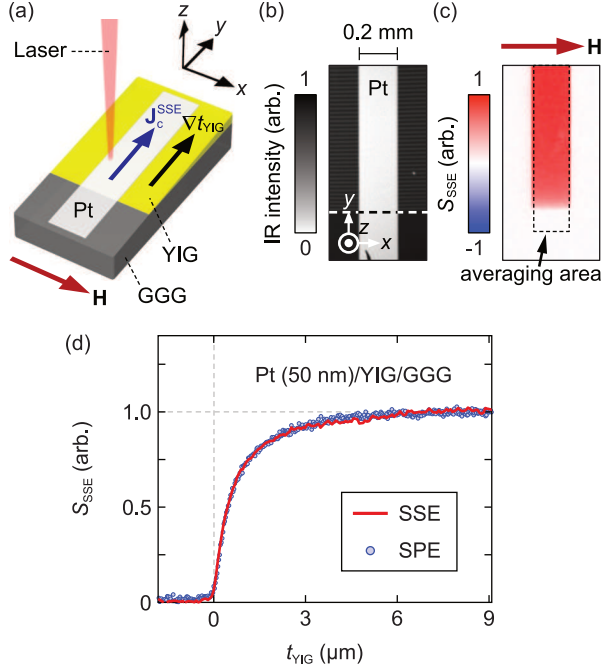}
\caption{(a) Schematic of the SSE measurement using a laser heating method. ${\bf{J}}_{\rm{c}}^{\rm{SSE}}$ is a charge current due to the inverse spin Hall effect induced by SSE. The thickness of the Pt film is $50$ nm, which is thick enough to prevent light transmission; the Pt layer is heated by the laser light. (b) Infrared light image of the sample. (c) The SSE signal $S_{\rm{SSE}}$ image induced by the laser heating. (d) $t_{\rm{YIG}}$ dependence of $S_{\rm{SSE}}$ and comparison with that of the temperature change induced by SPE. Here, we used an arbitrary unit for the SSE signal because the temperature gradient induced by the laser heating cannot be estimated experimentally. Nevertheless, the relative change of $S_{\rm{SSE}}$ is reliable because the heating of the Pt layer is uniform over the sample.}
\label{figure4}
\end{center}
\end{figure*}
%
%
%
%
%
%
%
To obtain the $t_{\rm{YIG}}$ dependence of SSE, the SSE signal was measured in the Pt/YIG/GGG sample by means of the thermoelectric imaging technique based on laser heating \cite{Iguchi2019,SSE4,SSE14,SSE29}. As shown in Fig. \ref{figure4}(a), the sample surface was irradiated by laser light with the wavelength of $1.3$ $\mu$m and the laser spot diameter of $5.2$ $\mu$m to generate a temperature gradient across the Pt/YIG interface. The local temperature gradient induces a spin current across the interface due to SSE. The spin current is then converted into a charge current via the inverse spin Hall effect in Pt \cite{SHE6}. To avoid the reduction of the spatial resolution of the SSE signal due to the thermal diffusion, we adopted a lock-in technique in the laser SSE measurement, where the laser intensity was modulated in a periodic square waveform with the frequency $f=5$ kHz and the thermopower signal $S_{1f}$ oscillating with the same frequency as that of the input laser was measured. The measurements at the high lock-in frequency realized high spatial resolution for the SSE signal because temperature broadening due to the heat diffusion is suppressed. Here, we defined the SSE signal $S_{\rm{SSE}}$ as $\left[S_{1f}(+50\ {\rm{mT}})-S_{1f}(-50\ {\rm{mT}})\right]/2$ to remove magnetic-field-independent background. By scanning the position of the laser spot on the sample, we visualized the spatial distribution of the SSE signal with high $t_{\rm{YIG}}$ resolution of $64$ nm.\par
Figure \ref{figure4}(c) shows the spatial distribution of $S_{\rm{SSE}}$ for the Pt/YIG/GGG sample. In response to the laser heating, the clear signal was observed to appear in the Pt film. The $t_{\rm{YIG}}$ dependence of the SSE signal is plotted in Fig. \ref{figure4}(d), where the $S_{\rm{SSE}}$ values were averaged along the $x$ direction in the area surrounded by the dotted line in Fig. \ref{figure4}(c). The SSE signal monotonically increases with increasing $t_{\rm{YIG}}$. Significantly, the $t_{\rm{YIG}}$ dependence of $S_{\rm{SSE}}$ shows the same behavior as that of the SPE-induced temperature modulation [Fig. \ref{figure4}(d)]. This result supports the reciprocity between SPE and SSE and strengthens our conclusion in the SPE measurement.\par
In the recent study on SSE in Ref. \onlinecite{SSE36}, Parakash {\it{et al.}} reported non-monotonical increase of the SSE signal with $t_{\rm{YIG}}$. Since the SSE signal takes a local maximum at $t_{\rm{YIG}} \sim l_{\rm{mp}}$, they estimated $l_{\rm{mp}}$ as $250$ nm from the maximum point. However, in our Pt/YIG samples, the SSE and SPE signals monotonically increase with increasing $t_{\rm{YIG}}$. These results suggest that $l_{\rm{mp}}$ is shorter than the $t_{\rm{YIG}}$ resolution, $64$ nm, in our experiments. The conclusion is consistent with the theoretical expectation of $l_{\rm{mp}}\sim 1$ nm \cite{SPEtheory2}.\par
%
%
%
%
%
%
%
%
%
%
In conclusion, we have discussed the length scale of the spin and heat transport by magnons in YIG by measuring the $t_{\rm{YIG}}$ dependence of SPE in the Pt/YIG sample. This measurement was realized by using the YIG film with the $t_{\rm{YIG}}$ gradient and the LIT method, which allow us to obtain the continuous $t_{\rm{YIG}}$ dependence of SPE in the single Pt/YIG sample. The experimental result is well reproduced by the phenomenological model based on the linear Boltzmann's theory for magnons referenced in Ref. \onlinecite{SPEtheory4} and $l_{\rm{m}}$ is estimated to be $3.9$ $\mu$m for our YIG sample. We also measured the $t_{\rm{YIG}}$ dependence of SSE. The SPE and SSE signals show the same behavior in the $t_{\rm{YIG}}$ dependence and monotonically increase as $t_{\rm{YIG}}$ increases. The monotonic increase implies that $l_{\rm{mp}}$ is shorter than $64$ nm for our YIG sample. These results give crucial information to understand the microscopic origin of the spin-heat conversion phenomena.\par
\begin{acknowledgments}
The authors thank R. Iguchi, T. Kikkawa, M. Matsuo, Y. Ohnuma, and G. E. W. Bauer for valuable discussions. This work was supported by CREST ``Creation of Innovative Core Technologies for Nano-enabled Thermal Management'' (JPMJCR17I1), PRESTO ``Phase Interfaces for Highly Efficient Energy Utilization'' (JPMJPR12C1), and ERATO ``Spin Quantum Rectification Project'' (JPMJER1402) from JST, Japan, Grant-in-Aid for Scientific Research (A) (JP15H02012) and Grant-in-Aid for Scientific Research on Innovative Area ``Nano Spin Conversion Science'' (JP26103005) from JSPS KAKENHI, Japan, the Inter-University Cooperative Research Program of the Institute for Materials Research (17K0005), Tohoku University, and NEC Corporation.
\end{acknowledgments}

\end{document}